\documentclass[conference]{IEEEtran}
\IEEEoverridecommandlockouts
\usepackage{cite}
\usepackage{amsmath,amssymb,amsfonts}
\usepackage{algorithmic}
\usepackage{graphicx}
\usepackage{textcomp}
\usepackage{xcolor}
\usepackage{bm}
\usepackage{float}
\usepackage{placeins}
\usepackage{makecell}

\usepackage{tikz}
\usepackage{pgfplots}
\usetikzlibrary{spy, arrows, patterns}
\pgfdeclarelayer{bg}    
\pgfsetlayers{bg,main}  

\usepackage{subcaption}          
\usepackage{booktabs}            
\usepackage{multirow}
\usepackage{rotating}           
\usepackage{numprint}           
\usepackage{hyperref}           

\definecolor{darkpastelpurple}{rgb}{0.59, 0.44, 0.84}
\definecolor{myred}{HTML}{E76F51}
\definecolor{myblue}{HTML}{376996}
\definecolor{mygreen}{HTML}{2A9D8F}
\definecolor{mypurple}{HTML}{822E81}
\definecolor{mywhite}{HTML}{f0dfbb}

\newcommand{\img}{\mathbf{x}}
\newcommand{\sysout}{\hat{\img}}
\newcommand{\latent}{\mathbf{y}}
\newcommand{\qlatent}{\hat{\latent}}
\newcommand{\synthparam}{\bm{\theta}}
\newcommand{\armparam}{\bm{\psi}}
\newcommand{\upparam}{\bm{\upsilon}}
\newcommand{\synth}{f_{\synthparam}}

\def\BibTeX{{\rm B\kern-.05em{\sc i\kern-.025em b}\kern-.08em
    T\kern-.1667em\lower.7ex\hbox{E}\kern-.125emX}}

\newcommand{\etal}{\textit{et al}., }
\newcommand{\ie}{\textit{i}.\textit{e}., }
\newcommand{\eg}{\textit{e}.\textit{g}.\ }
\newcommand{\shorteq}{
  \settowidth{\@tempdima}{-}
  \resizebox{\@tempdima}{\height}{=}
}

\begin{document}

\title{Low-complexity Overfitted Neural Image Codec}

\author{\IEEEauthorblockN{Thomas Leguay, Th\'eo Ladune, Pierrick Philippe, Gordon Clare, F\'elix Henry}
\IEEEauthorblockA{\textit{Orange Innovation}, France \\ \footnotesize \texttt{firstname.lastname@orange.com}}
\and
\IEEEauthorblockN{Olivier Déforges}
\IEEEauthorblockA{\textit{IETR}, France \\ \footnotesize \texttt{olivier.deforges@insa-rennes.fr}}
}

\IEEEpubid{\makebox[\columnwidth]{{\copyright}2023 IEEE \hfill} \hspace{\columnsep}\makebox[\columnwidth]{ }}

\maketitle

\begin{abstract}
We propose a neural image codec at reduced complexity which overfits the
decoder parameters to each input image. While autoencoders perform up
to a million multiplications per decoded pixel, the proposed approach
only requires \numprint{2300} multiplications per pixel. Albeit low-complexity,
the method rivals autoencoder performance and surpasses HEVC performance under
various coding conditions. Additional lightweight modules
and an improved training process provide a 14\% rate reduction with respect
to previous overfitted codecs, while offering a similar complexity. This
work is made open-source at \small{\url{https://orange-opensource.github.io/Cool-Chic/}}.
\end{abstract}


\section{Introduction}

In many use cases (TV, video on demand), videos are encoded once using a
dedicated device, while they are decoded many times on a variety of low-power
devices such as smartphones. Consequently, the complexity and energy consumption
are substantially constrained for the decoder. For decades, conventional codecs
(H.264/AVC \cite{avc}, H.265/HEVC \cite{hevc} and H.266/VVC \cite{vvc}) have
been designed with this constraint in mind. Each successive codec provides
enhanced compression performance while still offering a low decoder complexity.

Conventional codecs have many different possibilities for compressing a signal.
Each time a signal is to be compressed, the encoder assesses the different
options available and only selects the few most suited ones. Encoding is framed
as a discrete optimization problem, \ie a competition between all parameters of
all tools, selecting the best ones through a rate-distortion (RD) cost. These
tools and the compressed signal are sent to the decoder which simply applies the
selected tools on the compressed signal. As such, the decoder complexity remains
low at the cost of an expensive encoding process which optimizes an RD cost for
each signal. Successive conventional codecs have provided an ever-increasing
number of hand-crafted coding tools while maintaining the individual tool
complexity sufficiently small. This leads to better signal adaptation and
improved compression performance.
\newline

Learned codecs based on autoencoders (AE) \cite{balle, ChengSTK20, aivc, elic}
offer an alternative coding paradigm. Here, the codec no longer looks for the
optimal decoding tools each time a signal has to be compressed. Instead, it is
designed during an offline training stage and computes in a single shot a
compressed representation optimizing an \textit{average} RD cost. Thus, the
instance-wise online RD optimization performed by conventional codecs is
replaced by an average offline optimization.

\begin{figure}
    \pgfdeclarelayer{bg}    
    \pgfsetlayers{bg,main}  
    \centering
    \begin{tikzpicture}
        \begin{semilogxaxis}[
            grid= major,
            width=\linewidth,
            height=6.8cm,
            xlabel = {Complexity [MAC / decoded pixels] $\downarrow$},
            ylabel = {BD-rate vs. HEVC (HM) [\%] $\downarrow$} ,
            xmin = 100, xmax = 1000000, xlabel near ticks, minor x tick num=0,
            ymin = -20, ymax = 20, ylabel near ticks, minor y tick num=0, ytick distance={10},
            title style={yshift=-0.75ex},
            ylabel shift=-0.15cm,
            legend style={at={(0.,0.)}, anchor= south west}
        ]

            \addplot[thick, myblue, only marks, mark=square*, mark size=4pt] coordinates {
                (83000 , 12.837)      
                (363350, -14.459)     

            };
            \addlegendentry{\sf \small Autoencoders}

            \addplot[thick, myred, only marks, mark=*, mark size=4pt] coordinates {
                (710, 14.964)        
                (802, 0.401)         
                (2287, -7.122)       

                };
            \addlegendentry{\sf \small Overfitted}

            \node [myred, right, xshift=0.15cm] at (axis cs:710, 14.964){\sf \small COOL-CHIC \cite{cool-chic}};
            \node [myred, right, xshift=0.15cm] at (axis cs:802, 0.401){\sf \textbf{Ours (light)}};
            \node [myred, right, xshift=0.15cm] at (axis cs:2287, -7.122){\sf \textbf{Ours (main)}};
            \node [myblue, right, xshift=0.15cm] at (axis cs:83000  , 12.837){\sf \small Ballé \cite{balle}};
            \node [myblue, left, xshift=-0.2cm] at (axis cs:363350 , -14.459){\sf \small Cheng \cite{ChengSTK20}};
        \end{semilogxaxis}
    \end{tikzpicture}
    \caption{Rate savings versus HEVC (HM) on the CLIC 2020 professional validation
    set \cite{clic20pro}. Negative results mean that less rate is required to
    achieve the same quality as HEVC.}
    \label{fig:complexity-bdrate}
\end{figure}
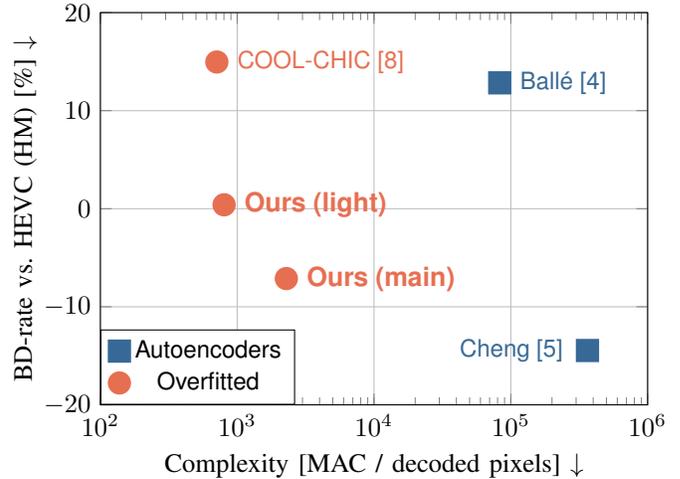

The JPEG-AI standardization effort shows that AE-based image codecs are
able to outperform conventional codecs \cite{jpeg-ai-cfp}. However, AE
decoders are often orders of magnitude more complex as they must offer good
performance on a wide variety of images. This results in prohibitive decoding
complexity \eg 800 kMAC / pixel (kilo multiplication-accumulation) for the
JPEG-AI development model \cite{jpeg-ai-complexity}. This might hinder the
practical usage of AE-based codecs.
\newline

\begin{figure*}
    \centering
    \includegraphics[width=0.8\textwidth]{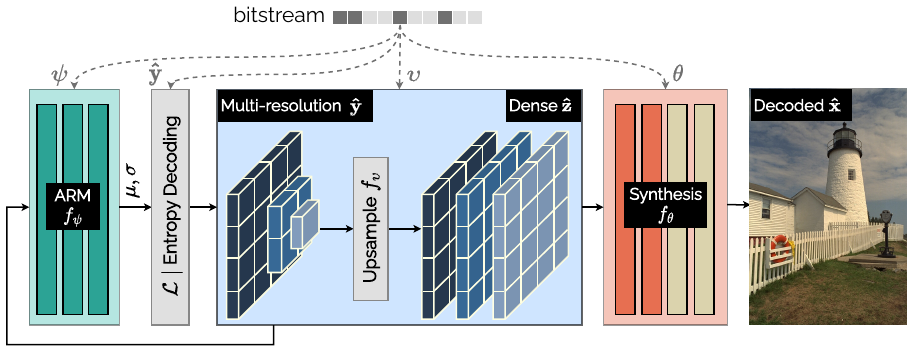}
    \caption{Decoding process of the proposed system. ARM stands for
    AutoRegressive Module and $\mathcal{L}$ is a Laplace distribution.}
    \label{fig:cool-chic}
\end{figure*}

Authors in \cite{shallow-decoder, CamposMDS19} propose to improve AE-based
codecs by reintroducing an instance-wise RD optimization. An initial solution
from an autoencoder is refined through gradient descent, overfitting the
autoencoder on the signal to compress. This suggests the possibility of using
overfitting to improve compression efficiency.

COOL-CHIC by Ladune \etal \cite{cool-chic} goes one step further, leveraging
Implicit Neural Representations \cite{coin} to design an overfitted image
codec without relying on an autoencoder. In Ladune \etal compressing an image
consists in overfitting a lightweight MLP (multilayer perceptron) decoder and a latent representation.
The overfitted MLP parameters and latent variables are then conveyed to the
receiver and used to reconstruct the image. This yields compression performance
on par with Ball\'e's autoencoder \cite{balle} for a decoder complexity a
hundred times smaller \ie of less than 1 kMAC / pixel.

By reintroducing an instance-wise RD optimization, this overfitting-based
approach provides compelling coding performance with a low decoder complexity.
Overfitting allows automatic learning of the adapted decoder for each image and
rate constraint. This avoids the hand-crafted design of the coding modes.
\newline

This paper enhances COOL-CHIC while maintaining a low
decoding complexity. Our contributions are as follows:
\begin{enumerate}
    \item A low-complexity adapted upsampling of the latent variables is
    proposed;
    \item Lightweight convolution layers are introduced into the original
    MLP-based decoder, improving its performance;
    \item Improved training. We show that considering the actual quantization
    instead of its relaxed version during the overfitting results in better
    performance.
\end{enumerate}
These contributions lead to a rate reduction of 14\%, outperforming HEVC for a
decoder complexity of 2.3 kMAC / pixel. In order to promote the design of
low-complexity codecs, this work is made open-source \cite{cool-chic-repo}.

\section{System overview}
 
This section presents the proposed decoding scheme based on \cite{cool-chic} and
illustrated in Fig. \ref{fig:cool-chic}. Let $\img \in \mathbb{N}^{C \times H
\times W}$ be an $H \times W$ image to compress, with $C$ color channels. The
proposed system is composed of three neural networks (NN): an auto-regressive
module (ARM) $f_{\armparam}$, an upsampler $f_{\upparam}$ and a synthesis
$f_{\synthparam}$. These NN are complemented with $\qlatent$, a set of $L$
two-dimensional multi-resolution discrete latent variables:
\begin{equation}
    \qlatent = \left\{ \qlatent_l \in \mathbb{Z}^{\frac{H}{2^l} \times \frac{W}{2^l}}, l \in 0, \ldots, L - 1 \right\}.
\end{equation}
The compressed representation of $\img$  comprises the NN weights
$\left\{\armparam, \synthparam, \upparam\right\}$ and the latent variable
$\qlatent$. The encoding stage optimizes these parameters while decoding
consists simply in applying the obtained parameters. 
\newline

\subsection{Decoding}

The first decoding step retrieves the NN parameters $\left\{\armparam,
\synthparam, \upparam\right\}$ from the bitstream following the method proposed
in \cite{cool-chic}. Then, each latent variable $\mathbf{\hat{y}}_l$ is entropy
decoded using a range coder driven by the ARM $f_{\armparam}$. The ARM models
the distribution of the $i$-th value from the $l$-th latent variable $y_{l,i}$
conditionaly to $\mathbf{y}_{l,<i}$, a set of already decoded values from the
same latent variable:
\begin{equation}
    y_{l,i} \sim \mathcal{L}(\mu_{l,i}, \sigma_{l,i}), \text{ where } 
    \mu_{l,i}, \sigma_{l,i} = f_{\armparam}(\mathbf{y}_{l,<i}).
\end{equation}
Then, the $L$ latents $\mathbf{\hat{y}}_l$ are upsampled to obtain a dense
latent representation $\hat{\mathbf{z}}$, with the spatial dimensions
of the image:
\begin{equation}
    \hat{\mathbf{z}} = f_{\upsilon}(\qlatent), \text{ with } \hat{\mathbf{z}} \in \mathbb{R}^{L \times H \times W}.
 \end{equation}
The synthesis transform is finally applied on the dense latent representation to compute
the decoded image $\sysout$:
\begin{equation}
   \sysout = \synth(\hat{\mathbf{z}}).
\end{equation}

\subsection{Encoding}

Encoding an image is achieved by overfitting the decoder parameters to determine
the optimal ones according to a rate-distortion (RD) objective:
\begin{equation}
    \left\{\qlatent, \armparam, \synthparam, \upparam\right\} = \arg\min D(\img,\sysout) + \lambda R(\qlatent),
    \label{eq:loss}
\end{equation}
with $D$ the mean-squared error, $R$ the rate (approximated by the entropy) and
$\lambda$ the Lagrange multiplier balancing rate and distortion. All the model
parameters are learned for a single image and are therefore adapted for its
content. 
\newline

Gradient descent is used to optimize the RD objective, requiring continuously
valued parameters. As such, the quantization of the latent representation
$\mathbf{\hat{y}} = Q(\mathbf{y})$ is modeled first as independent noise
addition \cite{DBLP:conf/iclr/BalleLS17} and then using a modified
Straight-Through Estimator \cite{STE-Q} (see Section \ref{sec:quantization}).
Once the encoding is finished, the NN parameters and the latent representation
are quantized and entropy coded to be sent efficiently to the decoder.

\begin{figure}
    \centering
    \begin{tikzpicture}
        \begin{axis}[
            axis lines=middle,      
            xmin=-2, xmax=2.5,      
            ymin=-0.5, ymax=1.49,     
            height=4.5cm, width=\linewidth,
            legend style={
                at={(1.05,1.15)},
                anchor=north east
            },
        ]
            \addplot[
                domain = -1.5:2.5,  
                samples=1000,       
                myred,              
                thick,              
                forget plot,        
            ] {-2 / 3 * x^3 + x^2 + 2 / 3 * x};
            \node[myred, above] at (axis cs:1.2, 1.1) {Interpolator $f(x)$};


            \addplot[
                ycomb,              
                myred,              
                thick,              
                mark=square*,       
                forget plot,        
            ] coordinates
            {
                (0.5, 0.5)
            };

            \node[myred, above] at (axis cs:0.3, 0.5) {$f(\tfrac{1}{2})$};

            \addplot[
                ycomb,              
                myblue,             
                thick,              
                mark=*,             
                forget plot,        
            ] coordinates
            {
                (-1, 1)
                (0 , 0)
                (1 , 1)
                (2 , 0)
            };

            \node[myblue, left] at (axis cs:-1, 1) {$s(-1)$};
            \node[myblue, above left] at (axis cs:0, 0) {$s(0)$};
            \node[myblue, left] at (axis cs:1, 1) {$s(1)$};
            \node[myblue, above right] at (axis cs:2, 0) {$s(2)$};

        \end{axis}
    \end{tikzpicture}
    \caption{Cubic interpolation of a discrete signal $s(n)$.}
    \label{fig:interpolation}
\end{figure}
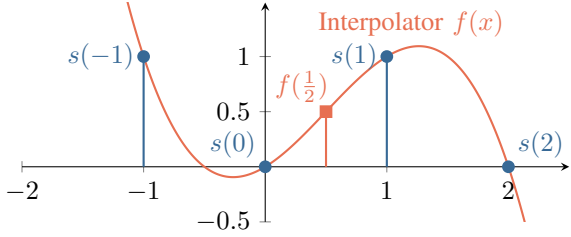

\section{Proposed improvements}

This section details the contributions of this paper:~designing an adapted
upsampling module, adding convolutional elements to the synthesis and better
considering the quantization during the training stage.

\subsection{Adapted upsampling}

The coding scheme relies on an upsampling step to obtain a dense representation
$\mathbf{\hat{z}}$ from the set of multi-resolution latent variables
$\mathbf{\hat{y}}$. In \cite{cool-chic}, a chained bicubic upsampling
\cite{bicubic-interpolation} by a factor of 2 is implemented. It is applied
multiple times  for each dimension to obtain the desired resolution. In this
section we recall that bicubic upsampling is equivalent to a convolution
operation. Here it serves as a proper initialization to learn a more adapted
upsampling.
\newline

For the sake of clarity, let us consider a one-dimensional discrete signal
$s\colon \mathbb{Z} \rightarrow \mathbb{R}$ whose value $s(n)$ is known for all
integers $n \in \mathbb{Z}$ (Fig. \ref{fig:interpolation}). In the system, the
upsampling step estimates the value $s(n~+~\frac{1}{2})$. Without loss of
generality, the case $n = 0$ is considered. Cubic upsampling defines an
interpolation function $f\colon [0, 1] \rightarrow \mathbb{R}$ as a third degree
polynomial:
\begin{equation}
    f(x) = \sum_{i=0}^{3} a_{i} x^{i} = \mathbf{a}^\mathsf{T} \mathbf{x}
    \text{ with } \mathbf{a} = \begin{bmatrix} a_0 \\ \vdots \\ a_3 \end{bmatrix}
    \text{, } \mathbf{x} = \begin{bmatrix} x^0 \\ \vdots \\ x^3 \end{bmatrix}.
    \label{eq:upsampling:interpolator}
\end{equation}

To obtain the polynomial coefficients $\mathbf{a}$,  the interpolator is
constrained to be equal to the actual signal for neighboring integers \ie $f(k)
= s(k), \forall k \in \left\{-1, 0, 1, 2\right\}$. Writing these conditions
using matrix notation yields:
\begin{equation}
    \mathbf{s} = \begin{bmatrix} s(-1) \\ s(0) \\ s(1) \\ s(2)\end{bmatrix} =
    \begin{bmatrix}
        1 & -1 & 1 & -1 \\
        1 &  0 & 0 &  0 \\
        1 &  1 & 1 &  1 \\
        1 &  2 & 4 &  8 \\
    \end{bmatrix}\begin{bmatrix} a_0 \\ a_1 \\ a_2 \\ a_3 \end{bmatrix}
    = \mathbf{B} \mathbf{a}.
    \label{eq:upsampling:condition}
\end{equation}

Combining equations \eqref{eq:upsampling:interpolator} and
\eqref{eq:upsampling:condition} allows to obtain the expression of the
interpolation function:
\begin{equation}
    f(x) = \left(\mathbf{B}^{-1}\mathbf{s}\right)^\mathsf{T}\mathbf{x}
    = \mathbf{s}^\mathsf{T}\left(\mathbf{B}^{-1}\right)^\mathsf{T}\mathbf{x}.
    \label{eq:upsampling:finalinterpolator}
\end{equation}

Since the upsampling always has a factor of 2, we are only concerned with
$f(\frac{1}{2})$ and $\mathbf{x}~=~\begin{bmatrix} 1 & \frac{1}{2} & \frac{1}{4}
& \frac{1}{8} \end{bmatrix}$. As such, eq.
\eqref{eq:upsampling:finalinterpolator} is written as the application of a
kernel $\upparam$ on the signal $\mathbf{s}$:
\begin{equation}
    f(\frac{1}{2}) = \upparam^{\mathsf{T}} \mathbf{s},
    \text{ with } \upparam = \frac{1}{16} \begin{bmatrix} -1 \\ 9 \\ 9 \\ -1 \end{bmatrix}.
\end{equation}

The same reasoning holds for all integers $n$ and for two-dimensional signals at
the expense of a two-dimensional (separable) kernel $\upparam$. Hence, computing
the upsampled value $f(n + \frac{1}{2})$ consists in convolving the original
signal $\mathbf{s}$ with $\upparam$.
\newline

\textbf{Contribution.} It is proposed to learn an adapted upsampling
$f_{\upparam}$. The encoder sets the desired filter properties (\eg,
cut-off frequencies or non-separability) to cope with the directional patterns
of the image. This gives the encoder more possibilities, allowing to
further optimize the rate-distortion trade-off. The upsampling kernel is
quantized and transmitted to the decoder similarly to the ARM and synthesis
parameters \cite{cool-chic}.

\begin{figure}
    \centering
    \includegraphics[width=\linewidth]{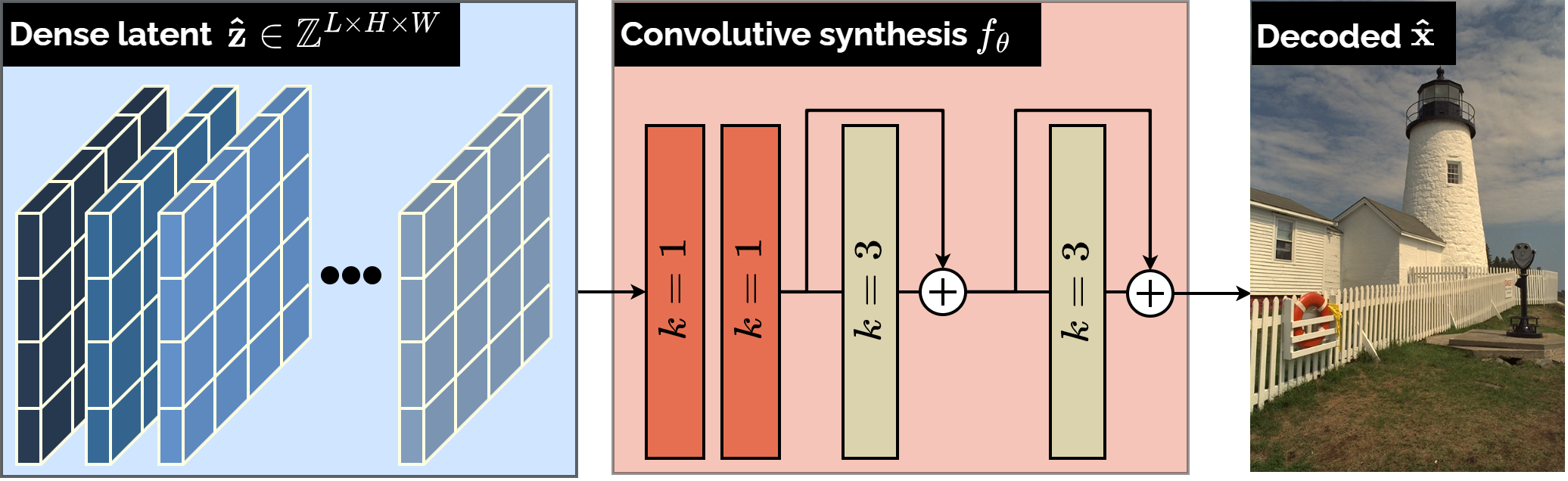}
    \caption{Convolution-based synthesis function $\synth$. The kernel size of each layer is denoted by $k$. More details in Table \ref{tab:config}.}
    \label{fig:postfilter}
\end{figure}

\begin{figure*}[ht]
    \centering
    \begin{subfigure}{0.48\linewidth}
        \begin{tikzpicture}
            \begin{axis}[
                grid= both,
                xlabel = {\small Rate [bpp] $\downarrow$},
                ylabel = {\small PSNR RGB [dB] $\uparrow$} ,
                xmin = 0, xmax = 1.0,
                ymin = 28, ymax = 42,
                ylabel near ticks,
                xlabel near ticks,
                width=\textwidth,
                height=7.5cm,
                xtick distance={0.2},
                ytick distance={2},
                minor y tick num=0,
                minor x tick num=0,
                legend style={at={(1.0,0.)}, anchor=south east},
            ]

                \addplot[dashed, thick, black, mark=none] table [x=rate_bpp,y=psnr_db] {data/clic20-pro-valid/vtm.txt};
                \addlegendentry{\small VVC (VTM)}

                \addplot[solid, thick, myblue, mark=*,mark size=1pt] table [x=rate_bpp,y=psnr_db] {data/clic20-pro-valid/cheng20.txt};
                \addlegendentry{\small Cheng et al. \cite{ChengSTK20}}

                \addplot[solid, ultra thick, mygreen, mark=square*,mark size=1pt] table [x=rate_bpp,y=psnr_db] {data/clic20-pro-valid/ours.txt};
                \addlegendentry{\small Ours}

                \addplot[solid, thick, black] table [x=rate_bpp,y=psnr_db] {data/clic20-pro-valid/hm.txt};
                \addlegendentry{\small HEVC (HM)}

                \addplot[solid, thick, mypurple, mark=*,mark size=1pt] table [x=rate_bpp,y=psnr_db] {data/clic20-pro-valid/balle18_hp.txt};
                \addlegendentry{\small Ballé et al. \cite{balle}}  

                \addplot[solid, thick, myred, mark=*,mark size=1pt] table [x=rate_bpp,y=psnr_db] {data/clic20-pro-valid/cool-chic.txt};
                \addlegendentry{\small COOL-CHIC \cite{cool-chic}}
            \end{axis}
        \end{tikzpicture}
        \caption{CLIC 2020 professional validation set.}
    \end{subfigure}
    \begin{subfigure}{0.48\linewidth}
        \begin{tikzpicture}
            \begin{axis}[
                grid= both,
                xlabel = {\small Rate [bpp] $\downarrow$},
                ylabel = {\small PSNR RGB [dB] $\uparrow$} ,
                xmin = 0, xmax = 1.0,
                ymin = 24, ymax = 38,
                ylabel near ticks,
                xlabel near ticks,
                width=\textwidth,
                height=7.5cm,
                xtick distance={0.2},
                ytick distance={2},
                minor y tick num=0,
                minor x tick num=0,
                legend style={at={(1.0,0.)}, anchor=south east},
            ]

                \addplot[dashed, thick, black, mark=none] table [x=rate_bpp,y=psnr_db] {data/kodak/vtm.txt};
                \addlegendentry{\small VVC (VTM)}

                \addplot[solid, thick, myblue, mark=*,mark size=1pt] table [x=rate_bpp,y=psnr_db] {data/kodak/cheng20.txt};
                \addlegendentry{\small Cheng et al. \cite{ChengSTK20}}

                \addplot[solid, ultra thick, mygreen, mark=square*,mark size=1pt] table [x=rate_bpp,y=psnr_db] {data/kodak/ours.txt};
                \addlegendentry{\small Ours}

                \addplot[solid, thick, black] table [x=rate_bpp,y=psnr_db] {data/kodak/hm.txt};
                \addlegendentry{\small HEVC (HM)}

                \addplot[solid, thick, mypurple, mark=*,mark size=1pt] table [x=rate_bpp,y=psnr_db] {data/kodak/balle18_hp.txt};
                \addlegendentry{\small Ballé et al. \cite{balle}}  

                \addplot[solid, thick, myred, mark=*,mark size=1pt] table [x=rate_bpp,y=psnr_db] {data/kodak/cool-chic.txt};
                \addlegendentry{\small COOL-CHIC \cite{cool-chic}}
            \end{axis}
        \end{tikzpicture}
        \caption{Kodak dataset.}
    \end{subfigure}
    \caption{Rate-distortion results on two datasets.}
    \label{fig:two-dataset}
\end{figure*}
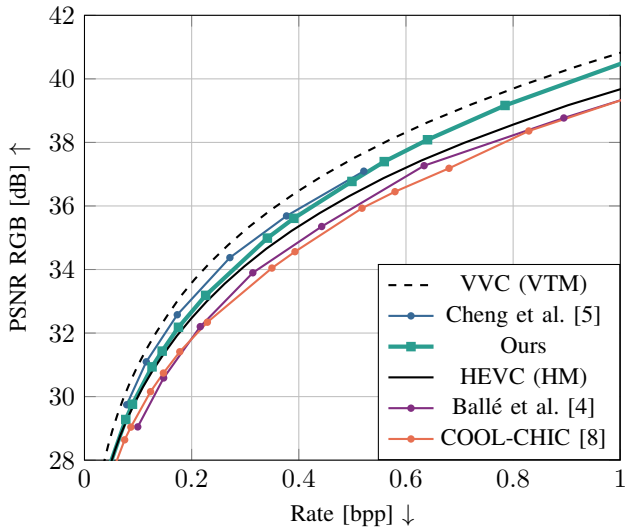
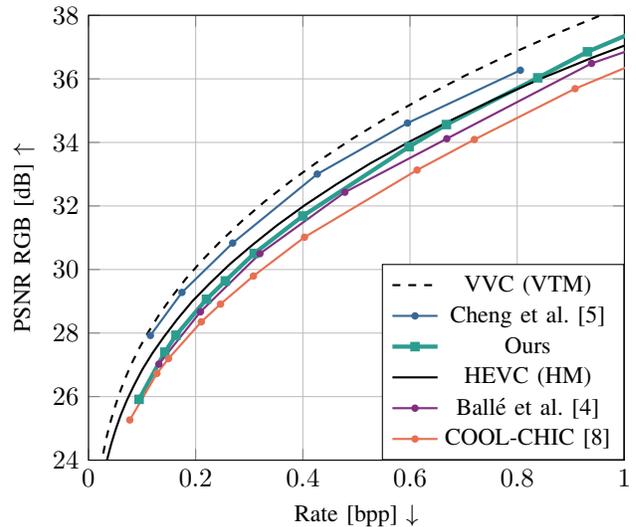

\subsection{Convolution-based synthesis}

The synthesis $f_{\synthparam}$ in \cite{cool-chic} is implemented as an MLP (\ie convolution layers with kernel of size 1). As such, it synthesizes
independently each pixel of the decoded image. 
\newline

\textbf{Contribution.} We introduce convolution layers with kernels of size 3 to
take advantage of neighboring latent values when synthesizing one pixel of the
decoded image. This new architecture is presented in Fig. \ref{fig:postfilter}.
As the synthesis operates under a strict complexity constraint, these
convolution layers are located at the end of the synthesis where they act as
residual post-filters on a 3-feature signal. Using kernel of size 3, such
convolution layer represents 81 MAC / decoded pixel. These layers are overfitted
alongside the whole system to obtain better rate-distortion performance.

\subsection{Quantization during training}
\label{sec:quantization}

\begin{figure}
    \centering
    \begin{tikzpicture}
        \begin{axis}[
            axis lines=middle,      
            xmin=-1, xmax=1.25,      
            xtick distance={1},
            ymin=-1.2, ymax=1.5,      
            height=5.4cm,
            width=\linewidth,
            legend style={
                at={(1.1,-0.05)},
                anchor=south east
            },
        ]
            \addplot[
                domain = -2:2.5,    
                samples=2000,       
                black,              
                thick,              
                forget plot,
            ] {round(x)};
            \node[black, above] at (axis cs:-0.87, -1) {$Q(x)$};



            \addplot[
                domain = -2:2.5,    
                samples=1000,       
                mygreen,            
                thick,              
                dashed,
                forget plot
            ] {1};
            \node[mygreen, above] at (axis cs:-0.88, 1) {STE};

            \addplot[
                domain = -2:2,      
                samples=1000,       
                myblue,              
                thick,              
                dashed,
                forget plot
            ] {0.1};
            \node[myblue, above] at (axis cs:-0.85, 0.1) {$\epsilon$-STE};

        \end{axis}
    \end{tikzpicture}
    \caption{Different approximations of the quantization derivative.}
    \label{fig:quant}

\end{figure}
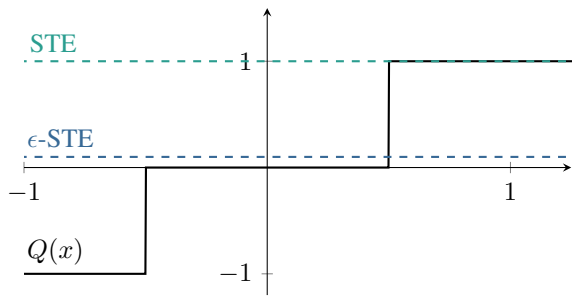

Quantization $Q(x) = \lfloor x \rceil$ is a key element of a lossy coding scheme
as it reduces the entropy of a signal. However, $\frac{\partial Q}{\partial x}(x)$
is null almost everywhere (see Fig. \ref{fig:quant}) preventing the usage of
gradient-based optimization methods. In the literature, quantization is often
modeled as a uniform noise addition during the training stage
\cite{DBLP:conf/iclr/BalleLS17}. In order to reduce the discrepancy between
training and inference some authors propose to switch to the actual quantization
at the end of the training \cite{guo2023learning}. In this case, a
straight-through estimator (STE) \cite{STE-Q} is used, setting $\frac{\partial
Q}{\partial x}(x) = 1$ manually in the backward pass. The same mechanism is
implemented in \cite{cool-chic}. 
\newline

\textbf{Contribution.} We argue that the STE is not the most suited gradient
estimator. Since $\frac{\partial Q}{\partial x}(x)$ is null almost everywhere,
setting a gradient close to zero is more consistent with the behavior of the
quantization function. We propose the $\epsilon$-STE which sets the derivative
to a small value \ie $\frac{\partial Q}{\partial x}(x) = \varepsilon$.
Empirically, $\varepsilon = 10^{-2}$ gives the best results when used alongside
the Adam optimization algorithm. This allows for a better optimization of the
RD cost during the encoding.

\section{Experiments}

\begin{figure*}
    \centering
    \begin{tikzpicture}
      \begin{axis}[
        width=0.93\textwidth,
        height=6cm,
        enlargelimits=0.05,
        ybar,
        bar width=0.13cm,
        ybar=-0.1cm,                           
        xmin=0, xmax=42, minor x tick num=0, xtick style={draw=none}, xmajorticks=false, xlabel near ticks,
        ymin=-35, ymax=35, ytick distance={10}, minor y tick num=0,
        ymajorgrids,
        axis x line=middle,
        axis y line=middle,
        every axis x label/.style={at={(current axis.right of origin)},anchor=west},
        every axis y label/.style={at={(current axis.above origin)},anchor=south},
        xlabel={Sequence},
        ylabel={BD-rate [\%] $\downarrow$},
    ]
        \begin{pgfonlayer}{bg}
            \draw [mygreen!20, fill=mygreen!20] (axis cs:0,0) rectangle (axis cs:42,-35);
            \draw [myred!20, fill=myred!20] (axis cs:0,0) rectangle (axis cs:42,35);
            \node[rotate=90] at (axis cs: 42.5,17.5) {Worse};
            \node[rotate=90] at (axis cs: 42.5,-17.5) {Better};
        \end{pgfonlayer}

        \addplot [myred, fill=myred] table[y=bd_rate, x=position] {data/clic20-pro-valid/sequence_wise_bd_rate_positive.csv};
        \addplot [mygreen, fill=mygreen] table[y=bd_rate, x=position] {data/clic20-pro-valid/sequence_wise_bd_rate_negative.csv};

        \node [below right] at (axis cs:18,-4) {\includegraphics[height=1.5cm]{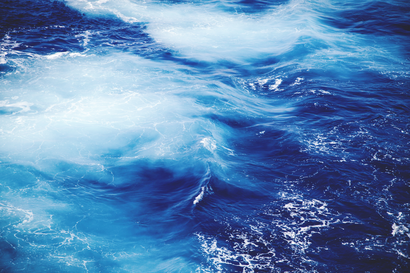}};
        \draw[->,-latex] (axis cs:19, -30) to[bend left=5] (axis cs:1.0, -30);
        \node [below] at (axis cs:29, -4) {\includegraphics[height=1.5cm]{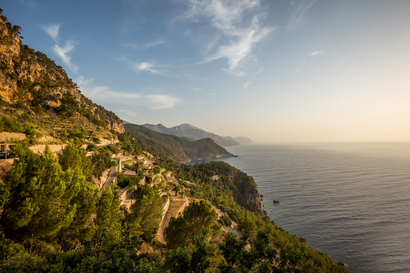}};
        \draw[->,-latex] (axis cs:27, -12) to[bend right=5] (axis cs:25, -1);

        \node [above] at (axis cs:28,4) {\includegraphics[height=1.5cm]{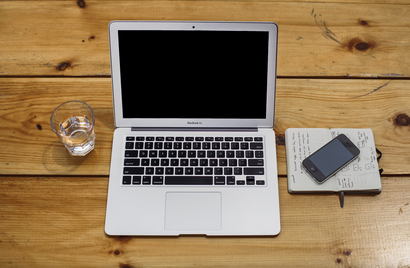}};
        \draw[->,-latex] (axis cs:31, 27) to[bend right=-5] (axis cs:41, 25);

    \end{axis}
    \end{tikzpicture}

    \caption{Sequence-wise BD-rate of the proposed system versus HEVC (HM) on
    the CLIC 2020 professional validation dataset. PSNR is measured in the RGB
    domain.}
    \label{fig:sequence-wise-rate}
\end{figure*}
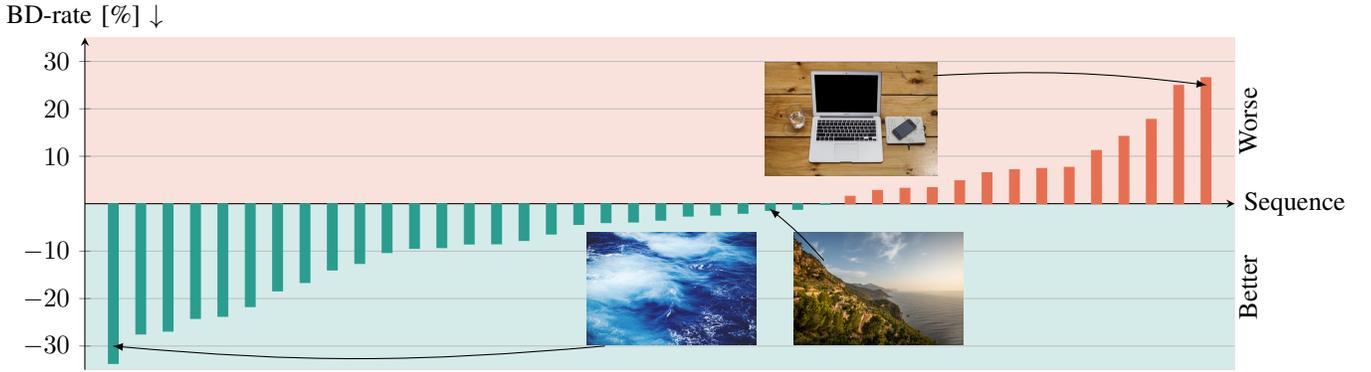

\subsection{Rate-distortion results}

\textbf{Experimental framework.} The proposed system is evaluated in two
configurations: light (worst case complexity of 0.8 kMAC / decoded pixel,
similar to \cite{cool-chic}) and main (worst case complexity of 2.3 kMAC /
decoded pixel). The details of both configurations are given in Table
\ref{tab:config}. The ARM uses respectively 12 or 24 spatial neighbours as
input. Both configurations have $L = 7$ latents and are optimized using the loss
function presented in eq. \eqref{eq:loss}. These models are compared to HEVC and
VVC through their reference implementations HM 16.20 and VTM 11.1  following the
conditions of \cite{jpeg-ai-cfp}. It is also compared to learned overfitted
codecs \cite{cool-chic} and autoencoder-based systems: Ballé \etal \cite{balle},
Cheng \etal \cite{ChengSTK20}. Performance of the codecs are expressed using the
BD-rate \cite{bdrate} \ie the relative rate required to achieve the same quality
(here the PSNR) than a reference codec (here HEVC).
\newline

\begin{table}[H]
    \centering
    \begin{tabular}{c|c|c|c}
        Config &       ARM $f_{\armparam}$ & Upsampling $f_{\upparam}$ & Synthesis $f_{\synthparam}$ \\
        \midrule
        \multirow{4}{*}{Main}        & 24 / 24 Linear              & TConv $k = 8\ s = 2$      & 7\hphantom{0}  / 40               Conv $k=1$      \\
                & 24 / 24 Linear              &                           & 40 / 3\hphantom{0}                Conv  $k=1$     \\
                & 24 / 2\hphantom{4} Linear   &                           & 3\hphantom{0} / 3\hphantom{0}     Conv $k=3$ \\
                &                             &                           & 3\hphantom{0} / 3\hphantom{0}     Conv $k=3$ \\
        \midrule
        \multirow{3}{*}{Light}                  & 12 / 12 Linear              & TConv $k = 8\ s = 2$      & 7\hphantom{0}  / 18               Conv $k=1$      \\
                & 12 / 12 Linear                &                             & 18 / 3\hphantom{0}                Conv  $k=1$     \\
                & 12 / 2\hphantom{4} Linear     &                             & 3\hphantom{0} / 3\hphantom{0}     Conv $k=3$ \\
        \end{tabular}
    \caption{Architecture of the proposed systems. $I$/$O$ indicates the number
    of input and output features. Each layer is followed by a ReLU, except the
    last one of each module. Kernel size and stride are denoted by $k$ and $s$
    respectively. TConv is a transpose convolution.}
    \label{tab:config}
\end{table}

\textbf{RGB performance.} Figure \ref{fig:complexity-bdrate} shows a
complexity-performance graph of both configurations on the CLIC 2020
professional validation set \cite{clic20pro}. The light system outperforms
\cite{cool-chic} while maintaining a similarly low complexity, proving the
relevance of the proposed improvements. Moreover, the light configuration is
competitive with HEVC (BD-rate: 0.4\%) and is more performant than Ballé \etal
at a significantly lower decoder complexity (0.8 vs. 83 kMAC / decoded pixel).
Finally, our main configuration outperforms HEVC (BD-rate: -7.1\%) with as few
as 2.3 kMAC / decoded pixel.

Figure \ref{fig:two-dataset} presents the rate-distortion curves of the main
configuration on the CLIC 2020 and Kodak \cite{kodak} datasets. On CLIC 2020,
the proposed model outperforms COOL-CHIC, Ballé \etal and HEVC on a wide range
of quality. At higher rates, it even comes close to Cheng \etal while having a
decoder-side complexity 100 times smaller. Similar albeit less favorable results
are observed on the Kodak dataset. Section \ref{sec:limitations} discusses the
causes of these less favorable results.
\newline

\textbf{Sequence-wise results.} Figure \ref{fig:sequence-wise-rate} presents the
sequence-wise BD-rate of the main system against HEVC. Out of the 41 images
of the CLIC dataset, 26 are better compressed with the main
configuration, highlighting its compelling results. The system offers
interesting performance on contents exhibiting fewer directional patterns (\eg
water) which are notoriously difficult for conventional codecs. This hints that
the system can be a interesting complement to conventional codecs.
\newline

\textbf{YUV420 results.} As a first step towards video coding, the proposed
system evaluated on the first frame of the MPEG Common Test Conditions videos
\cite{VVC_CTC}. Due to the YUV420 format, a nearest neighbour downscaling is
added after the synthesis. The performance against HEVC is presented in Table
\ref{tab:yuv}. It is often empirically noticed that PSNR YUV420 is less
favorable to learned codecs since conventional ones are designed with YUV420 in
mind. Here, this leads to worse compression performance of our system versus
HEVC. Yet, it still outperforms HEVC on specific contents such as the class F
corresponding to screen content.
\newline

\begin{table}[H]
    \centering
    \begin{tabular}{c | c c c c c | c}
        Class        & B & C & D & E & F & Average \\
        \midrule
        \makecell{Average \\ BD-rate [\%] $\downarrow$} & \textcolor{myred}{19.4} & \textcolor{myred}{8.0} & \textcolor{myred}{25.0} & \textcolor{myred}{34.4} & \textcolor{mygreen}{\textbf{-5.6}} & \textcolor{myred}{16.2}
    \end{tabular}
    \caption{Average BD-rate of the proposed main configuration versus HEVC.
    PSNR computed in the YUV420 domain.}
    \label{tab:yuv}
\end{table}

\textbf{Visual results.} Figure \ref{fig:visual_comparison} presents the same
image compressed at two rate targets by HEVC and the proposed system. At low
rate, HEVC exhibits blocking and ringing artifacts detrimental to the quality.
Since our system is not based on blocks, there is no such artifact. At higher
rates, both codecs are able to deliver high visual quality.

\newcommand{\subfigurewidth}{0.194\linewidth}  
\newcommand{\magnification}{6}
\newcommand{\spyon}{-0.75,0.25}
\newcommand{\zoomat}{1.74,-0.34}
\begin{figure*}[h]
   \centering
   \begin{subfigure}{\subfigurewidth}
      \begin{tikzpicture}[spy using outlines={myred,magnification=\magnification ,size=1.5cm}]
         \node {\includegraphics[width=\textwidth]{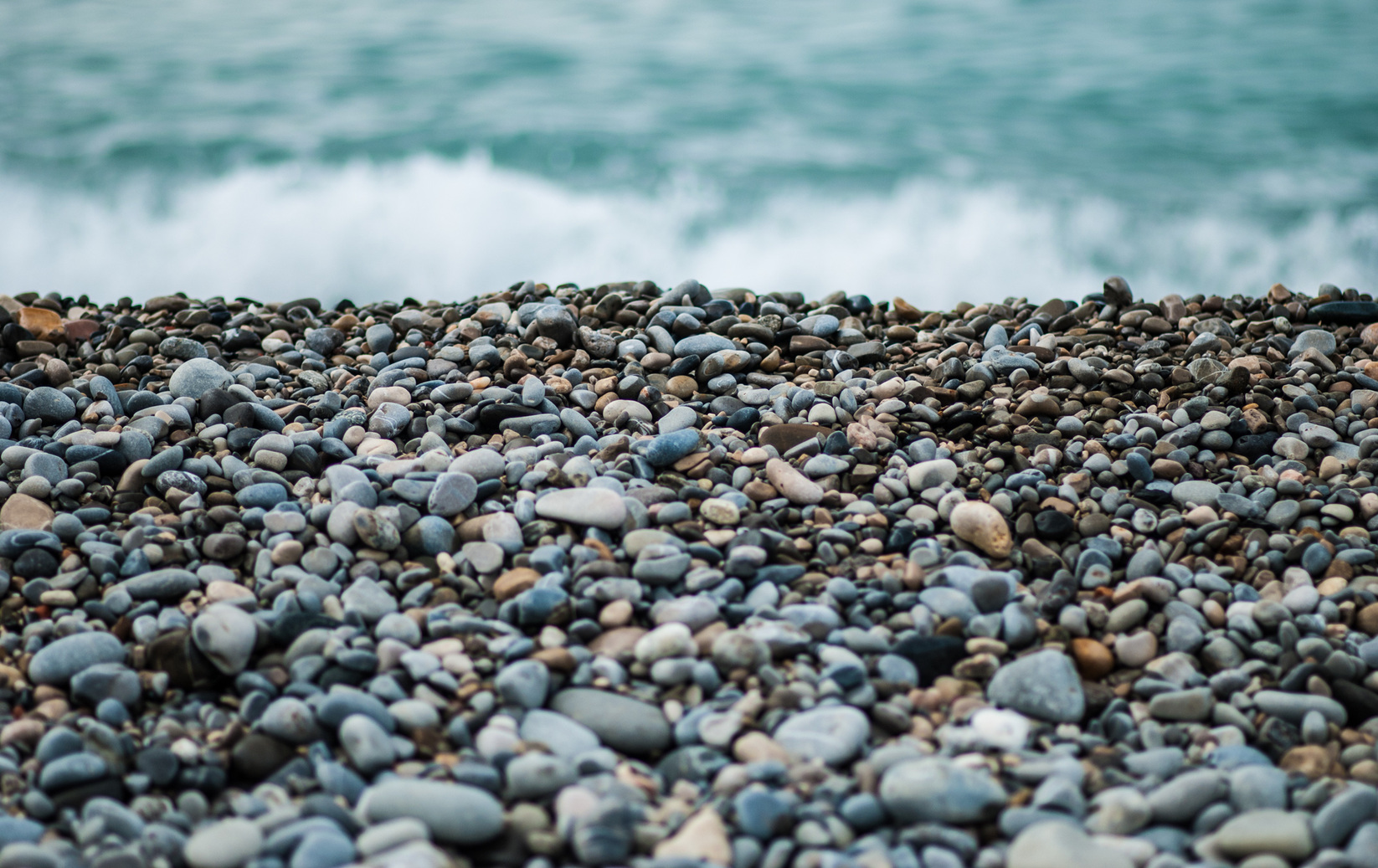}};
         \spy on (\spyon) in node [left] at (\zoomat);
      \end{tikzpicture}
      \caption{Original}
   \end{subfigure}
   \begin{subfigure}{\subfigurewidth}
    \begin{tikzpicture}[spy using outlines={myred,magnification=\magnification ,size=1.5cm}]
        \node {\includegraphics[width=\textwidth]{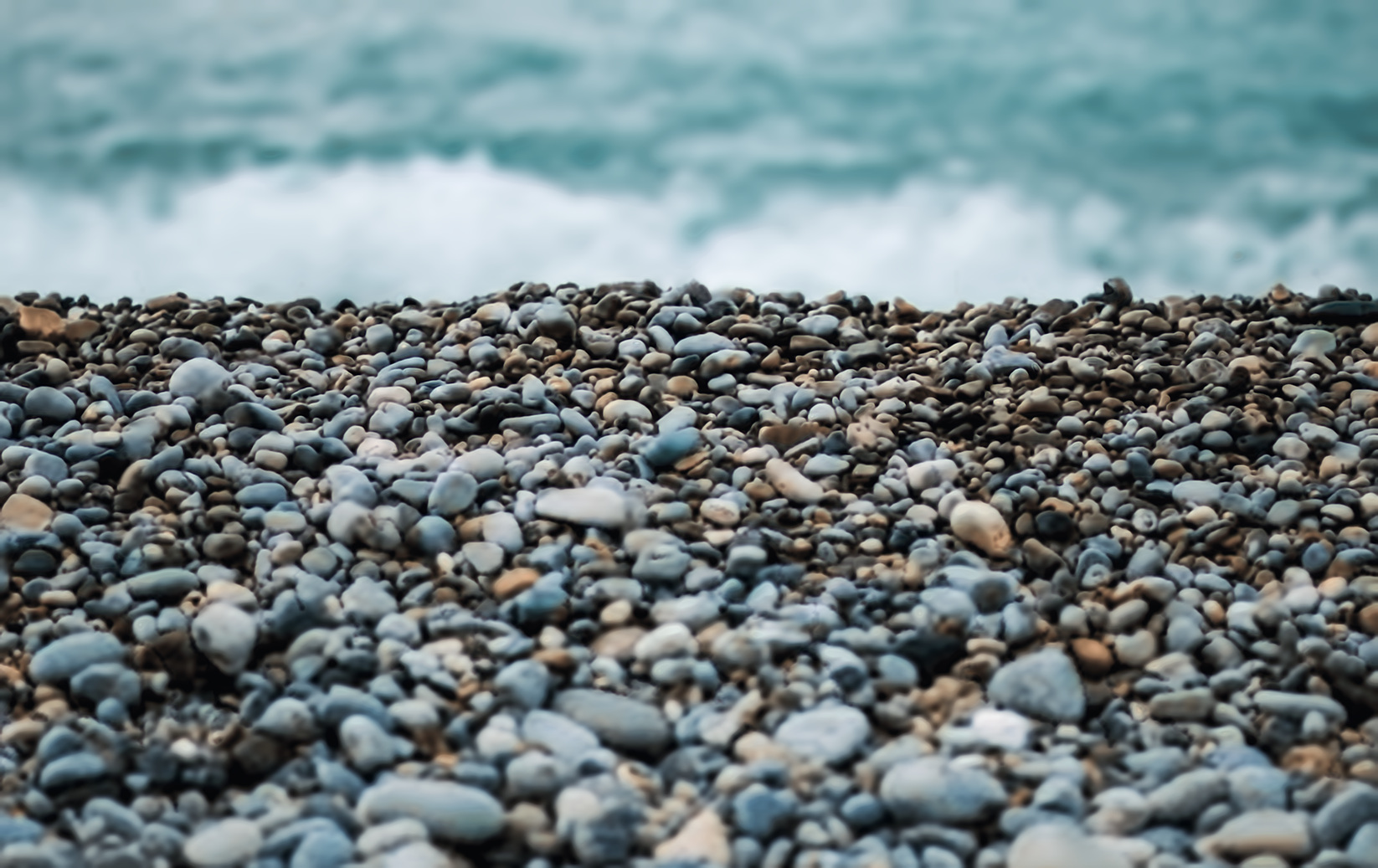}};
         \spy on (\spyon) in node [left] at (\zoomat);
      \end{tikzpicture}
      \caption{Ours 0.077 bpp}
   \end{subfigure}
   \begin{subfigure}{\subfigurewidth}
    \begin{tikzpicture}[spy using outlines={myred,magnification=\magnification ,size=1.5cm}]
        \node {\includegraphics[width=\textwidth]{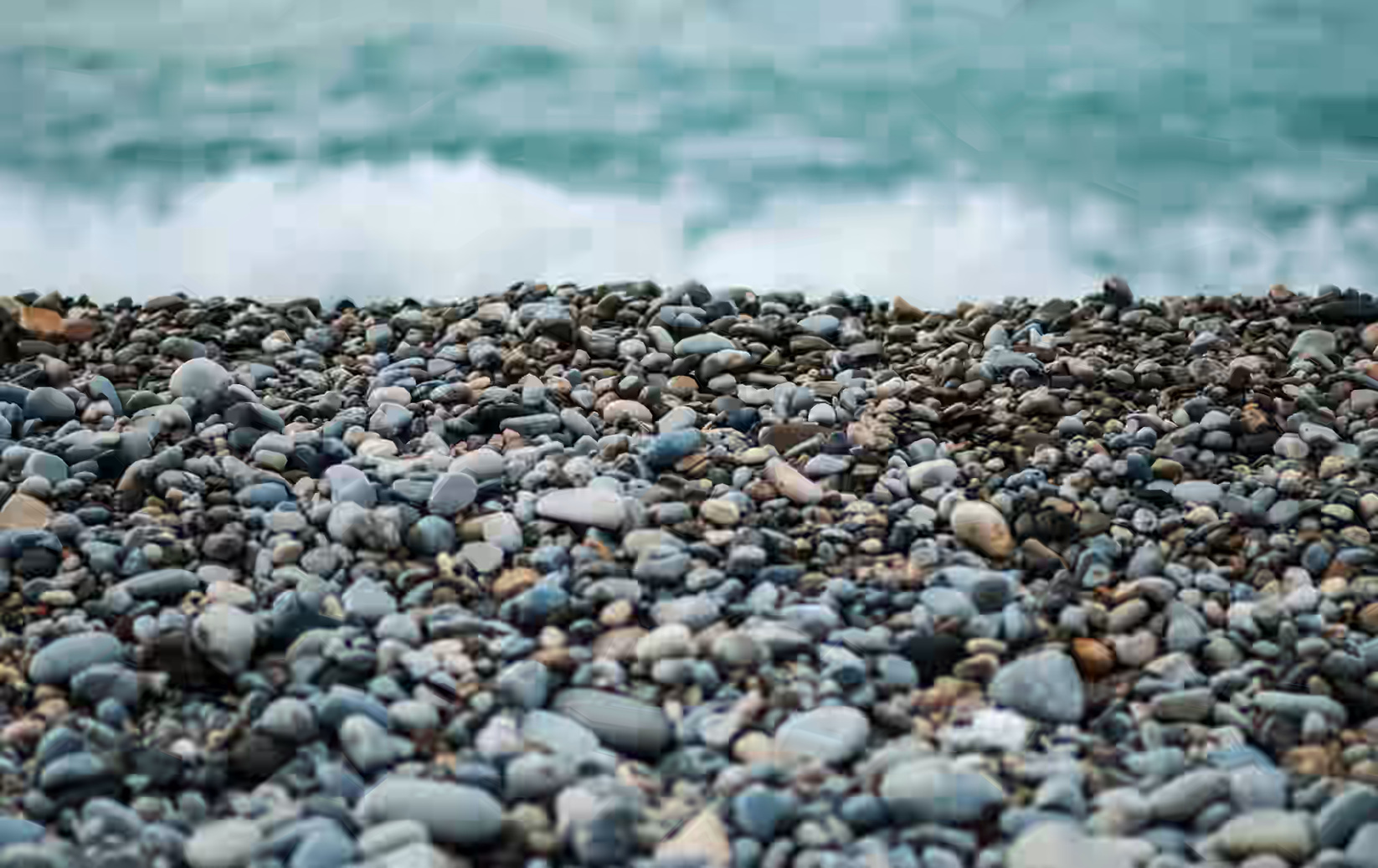}};
         \spy on (\spyon) in node [left] at (\zoomat);
      \end{tikzpicture}
      \caption{HEVC 0.074 bpp}
      \label{subfig:visualcomparison_low_rate}
   \end{subfigure}
   \begin{subfigure}{\subfigurewidth}
    \begin{tikzpicture}[spy using outlines={myred,magnification=\magnification ,size=1.5cm}]
        \node {\includegraphics[width=\textwidth]{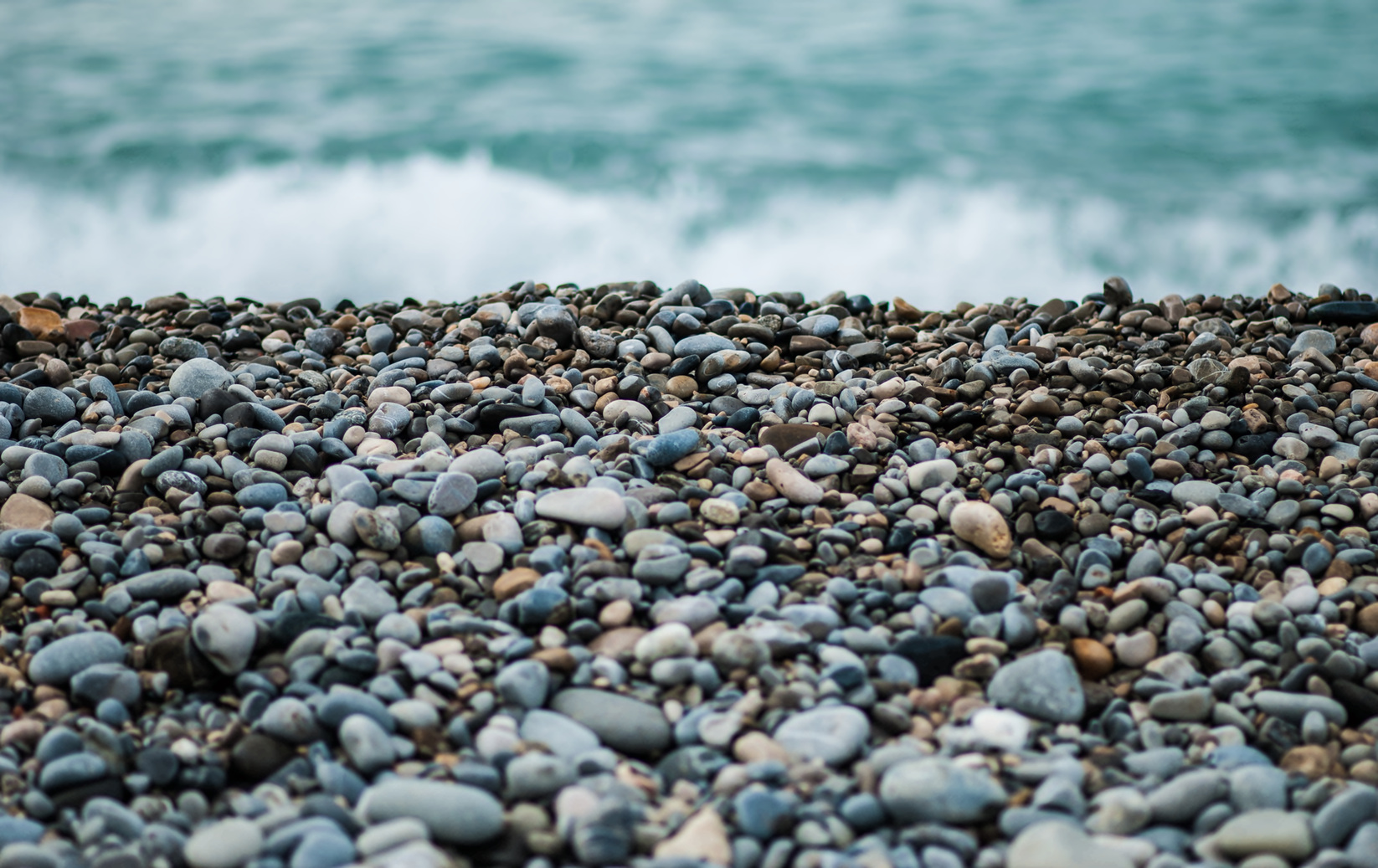}};
         \spy on (\spyon) in node [left] at (\zoomat);
      \end{tikzpicture}
      \caption{Ours 0.801 bpp}
   \end{subfigure}
   \begin{subfigure}{\subfigurewidth}
    \begin{tikzpicture}[spy using outlines={myred,magnification=\magnification ,size=1.5cm}]
        \node {\includegraphics[width=\textwidth]{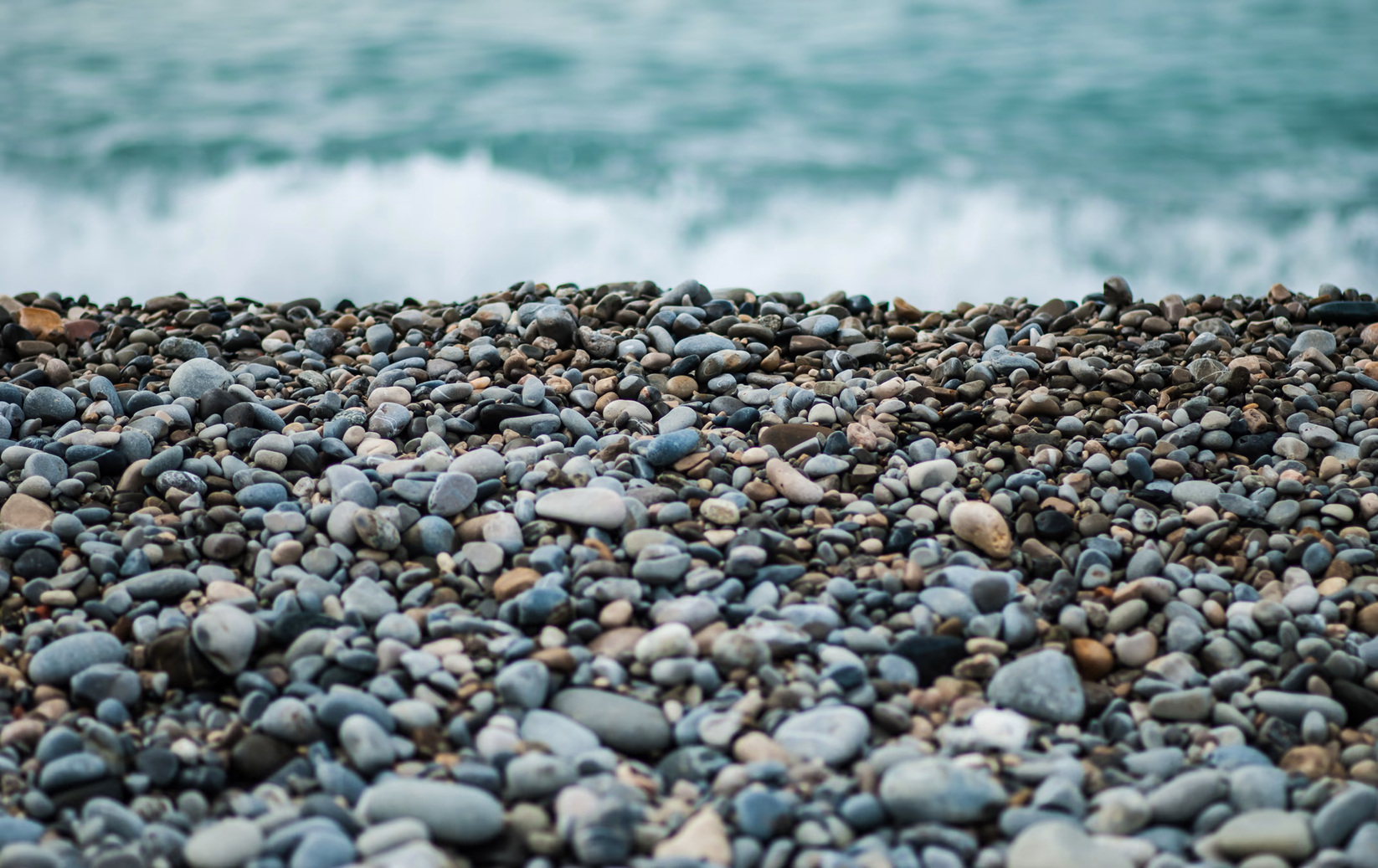}};
         \spy on (\spyon) in node [left] at (\zoomat);
      \end{tikzpicture}
      \caption{HEVC 0.813 bpp}
      \label{subfig:visualcomparison_high_rate}
   \end{subfigure}
   \caption{Comparison of HEVC and our system on CLIC 2020 image \textit{jeremy-cai-1174} at
   two different rates.}
   \label{fig:visual_comparison}
\end{figure*}

\subsection{Ablation study}

\begin{table}[ht]
    \centering
    \begin{tabular}{ccc | ccc | c}
        &&& \multicolumn{3}{c|}{\multirow{2}{*}{Complexity [kMAC / decoded pix]}}& \multirow{2}{*}{BD-rate vs.}\\
        &&& && & \\
        \begin{rotate}{90} \makecell{Conv. \\ synthesis} \end{rotate} & \begin{rotate}{90} \makecell{\\ Learned \\ upsample} \end{rotate} & \begin{rotate}{90} \makecell{\\ $\epsilon$-STE} \end{rotate}  & Synthesis  & Upsampling & Total & full model [\%] $\downarrow$\\
        \midrule
        \hfill & \hfill & \hfill & 0.8 & 0.03 & 2.5 & 14.0\\
        \checkmark & \hfill & \hfill & 0.6 & 0.03 & 2.2 & 9.6 \\
        \checkmark & \checkmark & \hfill & 0.6 & 0.1 & 2.3 & 6.0\\
        \checkmark & \checkmark & \checkmark & 0.6 & 0.1 & 2.3 & 0.0
    \end{tabular}
    \caption{Ablation study of the main configuration on CLIC 2020. ARM
    $f_{\armparam}$ remains identical for all tests and has a complexity of 1.6
    kMAC / decoded pixel. When disabled, $\epsilon$-STE is replaced by STE.}
    \label{tab:ablation}
\end{table}

Table \ref{tab:ablation} presents the BD-rate loss when disabling particular
modules. Removing all contributions corresponds to \cite{cool-chic}. The
ablation shows that each contribution increases performance while keeping the
overall complexity constant. Note that the best performance gain is due to the
$\epsilon$-STE during training \ie to a better optimization of the decoder. This
hints that better performance could be obtained without increasing the decoder
complexity by further improving the training stage.

\section{Limitations and future work}
\label{sec:limitations}

\begin{figure}
    \centering
    \begin{tikzpicture}
        \begin{axis}[
            grid= both,
            xlabel = {\small Total rate [bpp]},
            ylabel = {\small Rate share of NN parameters [\%]} ,
            xmin = 0., xmax = 1.0,
            ymin = 0, ymax = 40,
            ylabel near ticks,
            xlabel near ticks,
            width=\linewidth,
            height=6cm,
            xtick distance={0.25},
            ytick distance={10},
            minor y tick num=0,
            minor x tick num=0,
        ]

            \addplot[thick, smooth, dashed, myred, mark=*, mark options={solid}] table [x=rate_bpp,y=share_rate_mlp] {data/clic20-pro-valid/share_rate_nn.tsv};
            \addlegendentry{\small CLIC20}

            \addplot[thick, dashed, myblue, mark=square*, mark options={solid}] table [x=rate_bpp,y=share_rate_mlp] {data/kodak/share_rate_nn.tsv};
            \addlegendentry{\small Kodak}

        \end{axis}
    \end{tikzpicture}
    \caption{Rate allocated to the NN parameters at different bitrates}
    \label{fig:rate_network}
\end{figure}
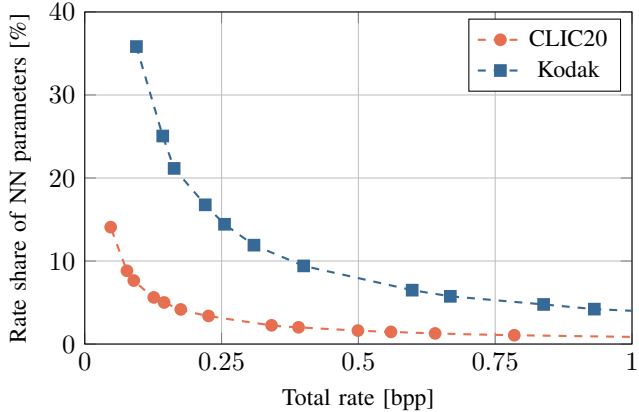

\textbf{NN parameters rate. } The proposed coding scheme assumes that the latent
variables represent most of the rate while the cost of sending the NN parameters
is neglectable. As such, the loss function does not take the network rate into
account, see eq. \eqref{eq:loss}. Yet, this assumption does not hold for all
types of images and rate targets. Figure \ref{fig:rate_network} presents the
share of rate allocated to the NN parameters. At lower rates or for
low-resolution images (Kodak), more than 10\% of the rate is dedicated to the NN
parameters, explaining the relatively worse results obtained for the Kodak
dataset. Future work should focus on reducing the rate of NN parameters by
considering techniques from the literature \eg pruning, tensor decomposition,
distillation \cite{network-compression}.
\newline

\textbf{Encoding. }The experimental results show that a better encoding \ie a
better optimization  (using $\epsilon$-STE) of the RD cost leads to significant
compression gains. Yet, this could be further refined by using more advanced
optimization techniques \cite{DBLP:conf/aaai/YaoGSMKM21} and more suited weights
initialization. Beside improving the compression efficiency
this would also reduce the encoding duration as it currently requires 10 to 60
minutes per image depending on the resolution. Note that a better implementation
would also drastically reduce the encoding time as demonstrated by Instant-NGP
\cite{mueller2022instant}.

\section{Conclusion}

This paper proposes a learned image coding scheme with a decoding complexity of
2.3 kMAC / pixel \ie a hundred times smaller than autoencoder-based codecs. This
lightweight codec offers up to 7\% rate reduction compared to modern
conventional codecs such as HEVC. It also significantly outperforms classical
autoencoders from Ballé \textit{et al.} This performance is achieved by
overfitting and conveying a neural decoder for each image. This paper refines
both the components of the decoder (upsampling, synthesis) and its overfitting
($\epsilon$-STE) leading to a rate reduction of 14\% compared to COOL-CHIC while
maintaining a low decoder-side complexity.

Future work should focus on improving the encoding process and better
compression of the neural network parameters in order to improve compression
efficiency without increasing the decoder-side complexity. 

{\small

\bibliographystyle{IEEEtran}
\bibliography{paper}
}

\end{document}